\documentstyle[aps,graphicx]{revtex}

\begin{document}
\draft
\preprint{RU97xx}

\twocolumn[\hsize\textwidth\columnwidth\hsize\csname @twocolumnfalse\endcsname

\title{Randomly Broken Nuclei and Disordered Systems}
\author{K.C. Chase \and P. Bhattacharyya \and A.Z. Mekjian}
\address{Department of Physics and Astronomy, Rutgers University \\
Piscataway, NJ 08855-0849}
\date{\today}

\maketitle

\begin{abstract}
Similarities between models of fragmenting nuclei and
disordered systems in condensed matter suggest corresponding methods.
Several theoretical models of fragmentation investigated in this
fashion show marked differences, indicating possible new methods for
distinguishing models using yield data.  Applying nuclear methods to
disordered systems also yields interesting results.
\end{abstract}

\vspace{0.3in}
]

\section{Introduction}
\label{sec:introduction}

The breakup of nuclei into clusters of various sizes 
in high energy heavy-ion collisions is currently being studied both
experimentally~\cite{Kreutz-P:1993npa,Hauger-JA:1996prl,Pochodzalla-J:1995prl,Wang-S:1995prl,Gilkes-ML:1994prl}
and
theoretically~\cite{Gross-DHE:1990rpp,Bondorf-JP:1995prep,Pan-JC:1995prc}.
The reasons for undertaking such studies are many, but
one is to determine whether a liquid-gas like phase
transition occurs at densities and temperatures away from the normal
conditions of the nucleus (achieved, albeit briefly, during
the collision).
By investigating the behavior of particular functions of the
fragmentation pattern, the character of the transition can
be determined, and the various critical point exponents can be measured.
For example, the EOS
collaboration~\cite{Gilkes-ML:1994prl,Elliott-JB:1994prc} studied a
set of moments introduced by Campi~\cite{Campi-X:1986jpa} to determine
the three critical exponents $\beta, \gamma, \tau$ and the critical
multiplicity.
Their results strongly suggest that nuclear fragmentation can be
understood in terms of a liquid-gas or possibly a percolative phase
transition.

Many non-nuclear systems also exhibit clustering.
For example, in condensed matter physics the phase space of disordered
systems~\cite{Derrida-B:1986jpa,Ziman-JM:1979-MoD}
such as spin
glasses~\cite{Mezard-M:1987-SGTaB,Chowdhury-D:1986-SGFS}
exhibits clustering of states of the system around energy minima.
As in the nuclear case, the sizes and distributions of the these
clusters are studied in an effort to better understand the
character of these systems and the nature of the disorder.
As such one is not surprised to discover a great degree of overlap in
studies of nuclear fragmentation and disordered
systems.  Differences in the approaches are considerable however.

This paper investigates in detail the correspondences
between fragmenting nuclear and disordered
systems, distinguishing the similarities and the differences in the approaches.
A similar investigation by Higgs~\cite{Higgs-PG:1995pre} discussed
such parallels between biological and disordered systems.
Our purpose is to enlarge the avenues of investigation in nuclear
fragmentation by examining areas where parallel methods of
analysis are likely to be fruitful.  
Such approaches have been
successful before, such as the application of percolation theory to
nuclear fragmentation~\cite{Bauer-W:1985plb,Campi-X:1986jpa} and our
preliminary results presented here and
elsewhere~\cite{Mekjian-AZ:1997pla} encourages such a
detailed look.  Our principal result, that all the models considered
can be realized as a sequential breaking of the interval, reconciles
the observation of these similarities to a distinct mechanism.

\section{Fragmentation Observables}
\label{sec:observables}

A nucleus of $A$ nucleons (or in general a system of $A$ objects)
can be partitioned into $n_{k}$ clusters of size $k$.  The total
number of clusters (in nuclear physics known as the multiplicity)
then is given by $m = \sum_{k} n_{k}$ and the total
number of nucleons $A = \sum_{k} k n_{k}$.
The set of $n_{k}$'s subject to $\sum_{k} k n_{k} = A$ determines the
partitions of $A$, with ${\mathbf{n}} = (n_{1},n_{2},\ldots n_{A})$
describing a particular partition or fragmentation.
Alternately, one can describe such a partition by
$\lambda = (\lambda_{1}, \ldots, \lambda_{m})$ where
$\lambda_{k}$ is the size of the $k$th cluster, $A = \sum_{k}
\lambda_{k}$, and the clusters are
ordered in some fashion, e.g. according to size
$\lambda_{1} \ge \lambda_{2} \ge \cdots \ge \lambda_{m}$
or for dynamical fragmentation by order of appearance.

Either $\mathbf{n}$ or $\lambda$ describes the clustering pattern, but
we are more interested in functions of the pattern which elucidates
the character of the fragmentation.  
The moments of a fragmentation pattern are often examined
for this reason, as they
distill the results of a complex fragmentation pattern into a
single number which can be simply interpreted.
We define the $s$th moment of a fragmentation pattern ${\mathbf{n}}$ by
\begin{equation}
Y_{s}({\mathbf{n}}) 
  = \sum_{k} k^{s} n_{k} \;.
    \label{eq:Ys-definition}
\end{equation}
The first two moments count the number of clusters and objects, i.e.
$Y_{0} = m$, $Y_{1} = A$.
The second moment $Y_{2} = Y$ indicates
the average size of the cluster ($Y/A$) a randomly chosen object
belongs to, and higher moments can be similarly understood.
In spin-glasses, $Y/A^{2}$ is known as the
participation ratio in Anderson localization~\cite{Anderson-PW:1958prev}.

Sometimes we wish to reduce these moments $Y_{s}$, eliminating any
contribution to them from the largest cluster, i.e. we define
the reduced $s$th moment as
\begin{equation}
M_{s}({\mathbf{n}}) 
  = Y_{s}({\mathbf{n}}) - k_{{\rm max}}^{s} \;.
    \label{eq:Ms-definition}
\end{equation}
where $k_{{\rm max}}$ is the size of the largest cluster.
Such a moment seems natural if we consider the system
as a liquid-gas system.  Then the largest cluster can be considered
the liquid phase, and such a reduction corresponds to considering the
contributions from the gas vapor only.  Alternately, percolation
theory would identify such a cluster with the incipient infinite
cluster, and the reduction would separate the system into percolating
and non-percolating regions.  In spin-glasses, such a reduction
removes the contribution of the extended state, the deepest valley in
the rugged energy landscape.

\section{Fragmentation Models}
\label{sec:models}

The discussion of observables considered only 
a single fragmentation event.
Physical systems produce such events in great numbers, but not
necessarily with a uniform distribution.
For the observables discussed above, once the distribution of events
is specified any question about the observable can in principle be
answered.
Clearly there is a physical mechanism which leads to
such a distribution, 
but it is possible that several mechanisms may yield
the same fragment distribution.
Analyzing ensemble averages or distributions on fragment observables
would then fail to distinguish between various mechanisms.
Therefore, one can specify a model by its underlying
fragment distribution without recourse to the mechanism which
generates it.
This section does this for a number of cases
applicable to nuclear fragmentation and disordered systems, and shows
that all these systems can be generated by a random breaking mechanism.

\subsection{Nuclear Systems}
\label{sec:models-nuclear}

When two heavy ions collide at intermediate energies, the nuclei
dissolve into a diverse mix of fragments.  The fragment distribution
is necessarily complex, but its general statistical features are
determined by simple thermodynamics.  For instance, the kinetic motion
of each fragment limits the available phase space, and so the
probability of seeing $m$ fragments is proportional to $x^{m}$,
where $x = V/\lambda_{T}^{3}$, the ratio of the available volume to
the thermal volume.  Internal excitations and other effects will
also enter, but in the large $T$ limit kinetic motion dominates,
predicting a large numbers of fragments.
Low temperature considerations likewise constrains our
fragment distribution.  At low $T$ a large fragment accompanied by some
very small fragments is expected.  The region between the two natural
limits having several intermediate mass fragments is
more difficult to describe and is the subject of current research.

One might believe that the distribution of fragments in the
intermediate region is not dominated by thermal factors, but rather by
the available phase space associated with the number of possible patterns.
Sobotka and Moretto~\cite{Sobotka-LG:1985prc} explored the simplest
such model in which every pattern is given equal probability.
Such a model shows an exponential fall-off in the fragment yield,
contrary to the experimentally observed power law.
One can generalize this model in many ways, the most obvious of 
which is to consider
a model where different cluster sizes would contribute different
amounts to the fragment distribution phase space, i.e. 
$\Pr({\mathbf{n}}) = (1/Z_{A}) \prod_{k} x_{k}^{n_{k}}$ so that 
$x_{k} = 1$ reduces to Sobotka and Moretto's model.
The partition function 
$Z_{A} = \sum_{{\mathbf{n}}} \prod_{k} x_{k}^{n_{k}}$ in this case can
be calculated using the recursion relation 
$Z_{A} = (1/A)\sum_{k>0} k \sum_{i>0} x_{k}^{i} Z_{A-ik}$ with
$Z_{0} = 1$.  

Such models have many interesting properties, but need 
some modifications to avoid problems associated with
the indistinguishability of clusters of the same size.  This
indistinguishability should reduce the size of the phase space
available to the fragments, or equivalently, reduce the probability
for a particular fragment distribution by
factors of $1/n_{k}!$ after the argument of Gibbs.  With this
consideration,  a simple and somewhat generic model of fragmentation
would be to take the probability of a particular fragmentation pattern
to be given by
\begin{equation}
\Pr({\mathbf{n}}) 
  = {1 \over Z_{A}} \prod_{k>0} {x_{k}^{n_{k}} \over n_{k}!}
    \label{eq:Gibbs-Pr}
\end{equation}
where the partition function or normalizing factor $Z_{A}$
can be obtained using the recursion
$Z_{A} = (1/A) \sum_{k} k x_{k} Z_{A-k}$ with $Z_{0} = 1$.
Such a model as applicable to nuclear fragmentation was explored
in~\cite{Chase-KC:1995prl,Chase-KC:1994prc1}, and took 
$x_{k} = x y^{k-1}/\beta_{k}$
where $x = V/\lambda_{T}^{3}$ arises from the thermal motion,
$y$ describes effects due to binding and
internal energies and $\beta_{k}$ allows different sized clusters to
contribute different amounts to the phase space.
Thermodynamic arguments discussed
in~\cite{Lee-SJ:1992prc2} allow $y$ to be expressed as
$y = \exp \left\{ {a_{V}/k_{B} T} +
(k_{B}T/\varepsilon_{0}) (T_{0}/(T + T_{0})) \right\}$
where $a_{V}$ is the binding energy per nucleon (i.e. 
it is assumed that $E_{B}(k) \approx a_{V}(k-1)$ for a cluster of size $k$),
$\varepsilon_{0}$ is the energy spacing between excited levels and
$T_{0}$ is a cutoff temperature for internal excitations.
The parameters $\beta_{k}$ can be taken as $k^{\tau}$ so as to
reproduce the observed cluster distribution, $\langle n_{k} \rangle
\sim k^{-\tau}$, where $\tau$ is a critical
exponent introduced by Fisher~\cite{Fisher-ME:1967phy}. 
For nuclear fragmentation $\tau \approx 2.2$~\cite{Gilkes-ML:1994prl}.
	
Several models with $\beta_{k} = k^{\tau}$ have been explored
by us and others.  For example, $\beta_{k} = 1$, called the linear chain
model, was extensively studied by Gross {\it et
al.}~\cite{Gross-DHE:1992adp1,Gross-DHE:1992npa,Jaqaman-HR:1992npa}.
Another example, $\beta_{k} = k$, was the first model investigated by one of
us (A.Z.M.)~\cite{Mekjian-AZ:1990prl}. We have investigated
$\beta_{k} = k^{\tau}$ with $\tau > 2$, in particular $\tau = 5/2$ more
recently~\cite{Chase-KC:1995prl}.  It has many features in common with
Bose-Einstein condensation and percolation theory; namely
below a certain value of $x = x_{c}$, 
a cluster forms containing a finite fraction of the mass of
the system even in the infinite $A$ limit.  This ``infinite'' cluster
can be likened to the accumulation of particles in the ground state in 
a Bose gas below $T=T_{c}$, or to the formation of the incipient infinite
cluster in percolating systems above $p=p_{c}$.

Related weights including more factors (some dynamical in nature)
have been studied by Gross {\it et al.}~\cite{Gross-DHE:1990rpp} and
Bondorf {\it et al.}~\cite{Bondorf-JP:1995prep} in describing
more detailed models of nuclear fragmentation.

\subsection{Disordered Systems}
\label{sec:models-disordered}

Disordered systems can also involve clustering, and in those cases the 
probability of seeing a particular clustering pattern is of interest.
In the case of nuclear fragmentation,
thermodynamic and phase space arguments were used to arrive at
the pattern distribution, i.e. Eq.~(\ref{eq:Gibbs-Pr}).  Here we 
arrive at a similar description by analyzing a simple but general
mechanism which generates disorder: the random sequential breaking of
an interval.  This model was studied extensively by Derrida and 
Flyvbjerg~\cite{Derrida-B:1987jdp,Derrida-B:1987jpa} and more recently
by Frontera {\it et al.}~\cite{Frontera-C:1995pre}.  Here we
generalize their mechanism in such a way as to encompass
Eq.~(\ref{eq:Gibbs-Pr}) and many other models.

The sequential breaking of an interval is simple to describe. In the
continuous case, the unit interval is sequentially broken into pieces
of sizes $W_{1}, W_{2}, \ldots$, each chosen randomly in some fashion.
Specifically, the sequence
$W_{1} = z_{1}$, $W_{2} = (1-z_{1})z_{2}$,
$W_{3} = (1-z_{1})(1-z_{2})z_{3}$ et cetera, is generated where
each $z_{k}$ is chosen from some probability distribution
$\rho_{k}(z)$.  In other words, at step $i$ one breaks off a randomly
chosen fraction 
$z_{i}$ of the remaining piece of size $1-W_{1}-\ldots W_{i-1}$.
Typically the probability distribution on $z_{i}$ is independent of
$i$, though that need not be the case.  
Derrida and Flyvbjerg studied the case where 
$\rho(z) = x(1-z)^{x-1}$~\cite{Derrida-B:1987jpa}.

Let us consider an equivalent, but discrete model.  In this model,
an interval of length $A$ is partitioned sequentially into
integral sized pieces.  Suppose a piece of size $k$ is broken off 
from the interval of size $A$ with probability 
\begin{equation}
\Pr(\lambda_{1} = k; A) 
  = {A-1 \choose k-1} 
    {(1)_{k-1} (x)_{A-k} \over (x+1)_{A-1}} \;,
    \label{eq:RBI-cut-Pr}
\end{equation}
where $(x)_{n} = x(x+1)\cdots(x+n-1)$ and $x > 0$ is a free parameter.
In the limit $A \rightarrow \infty$, $k/A \rightarrow W$, this
distribution converges to the continuous distribution mentioned above.

Applying this process repeatedly until the interval is exhausted,
the sequence $\lambda_{1}, \lambda_{2}, \ldots, \lambda_{m}$ is
generated.
The overall probability for this sequence appearing from this
process is the product
$\Pr(\lambda_{1}; A) \Pr(\lambda_{2}; A-\lambda_{1})
\Pr(\lambda_{3}; A-\lambda_{1}-\lambda_{2}) \cdots
\Pr(\lambda_{m}; A-\lambda_{1}-\ldots-\lambda_{m-1})$ and is
given by
\begin{eqnarray}
&& \lefteqn{\Pr(\lambda_{1}, \ldots, \lambda_{m})
  = {x^{m}} {A+x-1 \choose A}^{-1}} 
    \nonumber \\
 && \hspace{0.3in} \times {1 \over A(A-\lambda_{1}) \cdots
    (A-\lambda_{1}-\ldots-\lambda_{m-1})} \;.
    \label{eq:RBI-Pr-lambda}
\end{eqnarray}
By summing over all possible orders of such a sequence, the
probability
$\Pr({\mathbf{n}}) = \sum_{{\lambda} \mapsto {\mathbf{n}}} \Pr(\lambda_{1}, \ldots, \lambda_{m})$ 
that a particular unordered clustering pattern appears is determined,
and is given by
\begin{eqnarray}
\Pr({\mathbf{n}})
  = {A+x-1 \choose A}^{-1} \prod_{k} {1 \over n_{k}!}
    \left( {x \over k} \right)^{n_{k}}
    \label{eq:RBI-Pr-nk}
\end{eqnarray}
where we have applied Eq.~(\ref{eq:perm-identity}).
This is in fact the distribution given by Eq.~(\ref{eq:Gibbs-Pr})
with $x_{k} = x/k$.  More generally, the sequential breaking 
process where the
probability of breaking off a fragment of size $k$ from an interval of
size $A$ is given by $\Pr(k; A) = (k x_{k}/A) Z_{A-k}/Z_{A}$ yields
Eq.~(\ref{eq:Gibbs-Pr}) as the fragmentation pattern.
This gives a possible mechanism for obtaining such a fragment
distribution, but it is not unique.  Another mechanism based on Markov
chains was explored by us before~\cite{Chase-KC:1994prc1}.

By modifying the random breaking mechanism we can obtain the
other model discussed in the previous section (the equipartition
model) as well.
Suppose one allows multiple clusters of the same
size to be broken off instead of just one.  For example, take the
probability that $m$ clusters of size $k$ are broken from an interval
of length $n$ as $\Pr(k, m; A) = (k x_{k}^{m}/A) (Z_{A-mk}/Z_{A})$.
Then the distribution of cluster sizes can be shown to be given by 
$\Pr({\mathbf{n}}) = (1/Z_{A}) \prod_{k} x_{k}^{n_{k}}$.
This is in fact a generalization of a well-known method for
generating equiprobable partitions introduced by Nijenhuis and
Wilf~\cite{Nijenhuis-A:1978-combin-algorithms}.

Having obtained the models of the previous section as cases of this
disordered system mechanism, we are encouraged to believe that other
disordered systems can be likewise described in the context of
randomly breaking the interval.   Indeed, two other models considered
by Derrida and Flyvbjerg fit nicely into this mechanism.

Consider the quenched random map
model~\cite{Derrida-B:1987jdp,Donnelly-PJ:1991aap,Derrida-B:1988jpa}.
The $A^A$ maps or 
functions of $A$ points onto themselves are chosen uniformly at
random, and their cluster structure is determined by the number of
points in each basin of attraction, where a basin is defined as a
limit cycle of the iterated action of the map.  
For example, the map 
$1 \mapsto 2, 2 \mapsto 1, 3 \mapsto 2, 4 \mapsto 3$ has a single
limit cycle ($1 \mapsto 2 \mapsto 1 \ldots$) which all the points
eventually enter, so that the cluster structure is just one single
cluster of size four.  Such a simple model has
interesting properties, and its cluster pattern is given by
Eq.~(\ref{eq:Gibbs-Pr}) with $x_{k} = \sum_{j=0}^{k-1} k^{j-1}/j!
\approx e^{k}/2k$.
This can be inferred from the fact that the probability that the map
is indecomposable is known and would be given by $x_{k}/Z_{k}$ where
$Z_{k} = k^{k}/k!$ by simple counting.   
Interpreting Eq.~(\ref{eq:Gibbs-Pr}) in terms of randomly breaking an
interval has already been accomplished, so we see that the random map
model is a special case of randomly breaking the interval.
The asymptotic limit can be seen by
considering the elements in the sum as terms in the Poisson
distribution, $e^{-x} x^{j}/j!$ with $x=k$.
The factor $e^{k}$ does not affect the cluster
distributions, so that the model asymptotically converges to
$x_{k} = 1/(2k)$, and hence is included in the
original distribution considered by Derrida and
Flyvbjerg~\cite{Derrida-B:1987jdp} $\rho(z) = x(1-z)^{x-1}$ with
$x=1/2$.

Secondly, consider what happens when the sequential pieces are chosen
from a changing distribution.
For example, suppose the $i+1$st piece is chosen from the interval of
size $A_{i} = A-\lambda_{1}-\ldots-\lambda_{i}, A_{0} = A$ with
probability
\begin{equation}
\Pr(k; A_{i}) 
  = {A_{i}-1 \choose k-1} 
    {(\gamma)_{k-1} (x+i(1-\gamma))_{A_{i}-k} \over 
     (x+\gamma+i(1-\gamma))_{A_{i}-1}} \;.
    \label{eq:PM-cut-Pr}
\end{equation}
In this case the sequence of pieces is generated with probability
\begin{eqnarray}
&& \lefteqn{\Pr(\lambda_{1}, \ldots, \lambda_{m})
  = A! {(x)_{(m-1; 1-\gamma)} \over (x+\gamma)_{A-1}}}
    \nonumber \\
 && \hspace{0.3in} \times 
    {1 \over A(A-\lambda_{1})(A-\lambda_{1}-\lambda_{2})\cdots}     
    \prod_{k} {(\gamma)_{\lambda_{k}-1} \over (\lambda_{k}-1)!} \;,
    \label{eq:PM-Pr-lambda}
\end{eqnarray}
where $x_{(n;\alpha)}$ is a generalized Pochhammer symbol
\[ x_{(n;\alpha)} = \left \{ 
\begin{array}{ll} 
  1 & \mbox{for $n = 0$} \\ 
  x(x+\alpha)\cdots(x+(n-1)\alpha) & \mbox{for $n > 0$} 
\end{array} \right.\]
Summing over all the possible orders of the $\lambda$'s,
gives the probability for the unordered fragmentation pattern
\begin{equation}
\Pr({\mathbf{n}}) 
  = A! {(x)_{(m-1;1-\gamma)} \over (x+\gamma)_{A-1}}
    \prod_{k} \frac{1}{n_{k}!} \left(
    \frac {(\gamma)_{k-1}}{k!} \right)^{n_{k}} \;.
    \label{eq:PM-Pr}
\end{equation}
This is clear since the only term that depends on the order of
the fragments is $1/(A(A-\lambda_{1})\cdots)$; the other terms are
invariant under permuting the order of the $\lambda$'s, so that we can
apply Eq.~(\ref{eq:perm-identity}) directly.

This distribution pattern has been studied in a statistics
context by J. Pitman~\cite{Pitman-J:1995ptrf}, but what we would like
to emphasize here is that in the limit $x+\gamma = 1$, this model
reproduces asymptotically all of the clustering features
of a Sherrington-Kirkpatrick
spin-glass~\cite{Sherrington-D:1975prl,Derrida-B:1980prl,Derrida-B:1987jpa}.
Unlike the other disordered systems, we cannot prove this result, but
the evidence (discussed below) is compelling enough that we consider
Eq.~(\ref{eq:PM-Pr}) with $x+\gamma = 1$ as a model of a spin-glass.
Why the changing distributions produces the patterns seen in a
spin-glass is perhaps unexpected and should be explored further.

\section{Traditional Analysis of Models}
\label{sec:traditional-analysis}

Now that we have a description of the fragment distributions we are in
a position to make definitive predictions.
Since nuclear and disordered systems have spawned separate methods of
analysis, we begin by briefly considering some features of these models
from their traditional points of view.

\subsection{Nuclear Systems}
\label{sec:traditional-nuclear}

Fragmenting nuclear systems are traditionally analyzed by
studying expectation values of the various observables as functions of
thermodynamic variables.  Some of the important observables are the
multiplicity $m$ and its variance, the size of the largest cluster
$k_{{\rm max}}$, and the reduced moment $M_{2}$.  They are important
because of their expected behavior in the region of a critical point
phase transition.  For example, from percolation theory we know that
the reduced moments are functions of the critical exponents and
$p - p_{c}$ in the region of the critical
point~\cite{Elliott-JB:1994prc}, i.e.
\begin{eqnarray}
\langle M_{0} \rangle 
  & = & \langle m \rangle - 1
   \sim |p-p_{c}|^{2-\alpha} 
        \nonumber \\
\langle M_{1} \rangle 
  & = & A - \langle k_{\rm max} \rangle 
   \sim |p-p_{c}|^{\beta} 
        \\
\langle M_{2} \rangle 
  & = & \langle Y_{2} \rangle - \langle k_{\rm max}^{2} \rangle
   \sim |p-p_{c}|^{-\gamma}
        \nonumber
\end{eqnarray}
The EOS collaboration~\cite{Gilkes-ML:1994prl} has argued that both
nuclear fragmentation and percolation theory can be analyzed in a
similar fashion, extending the observations and results of
Campi~\cite{Campi-X:1986jpa,Campi-X:1992npa}.
Thus from observing the behavior of the moments the
critical point and exponents can be deduced, and
a number of papers in the
literature have studied this in a variety of
theoretical~\cite{Campi-X:1992npa,Chase-KC:1995prc,Jaqaman-HR:1992npa} 
and experimental~\cite{Kreutz-P:1993npa,Gilkes-ML:1994prl} systems.

In the model given by Eq.~(\ref{eq:Gibbs-Pr}),
these quantities can be computed from their definitions and by
utilizing the following 
results
\begin{eqnarray}
\langle n_{k} \rangle 
  & = & x_{k} {Z_{A-k} \over Z_{A}} \;,
        \\
\langle n_{j}(n_{k}-\delta_{jk}) \rangle
  & = & x_{j} x_{k} {Z_{A-j-k} \over Z_{A}} \;.
\end{eqnarray}
Computation of the reduced moments additionally requires that various
expectation values of $k_{{\rm max}}$ be determined.  This is most
easily done by expressing the expectation values in terms of the
probability distribution on $k_{{\rm max}}$.
The probability that a random fragmentation of nucleons has a cluster
of size $k_{{\rm max}}$ as its largest cluster is given by
$\Pr(k_{\rm max}) = \Delta Z_{A}(x_{1}, \ldots,
x_{k_{{\rm max}}})/Z_{A}(x_{1},\ldots,x_{n})$ where
$\Delta Z_{A}(x_{1}, \ldots, x_{k_{{\rm max}}}) = 
Z_{A}(x_{1},\ldots,x_{k_{\rm max}},0,\ldots) - 
Z_{A}(x_{1},\ldots,x_{k_{\rm max}-1},0,\ldots)$.  When
$k_{\rm max} > A/2$ a simpler expression can be used.
Since any cluster of size $k>A/2$ is automatically the largest one can
show that $\Delta Z_{A} = x_{k} Z_{A-k}$, so that
$\Pr(k_{\rm max} = k) = x_{k} Z_{A-k}({\mathbf{x}})/Z_{A}({\mathbf{x}})$
when $k > A/2$.

\begin{figure}
\caption{The expected value (a) and variance (b) of the multiplicity
and the expected values of the reduced second
moment (c) and the largest cluster (d)
for the model given by Eq.~(\ref{eq:Gibbs-Pr}) 
with $x_{k} = x/k^{\tau}$, $\tau = 2.5$ (solid line), 
$\tau = 3.0$ (dashed line).}
\label{fig:Exp-nuclear}
\centerline{\includegraphics[bb = 43 20 592 746, angle = -90,
            width = 6.0in]{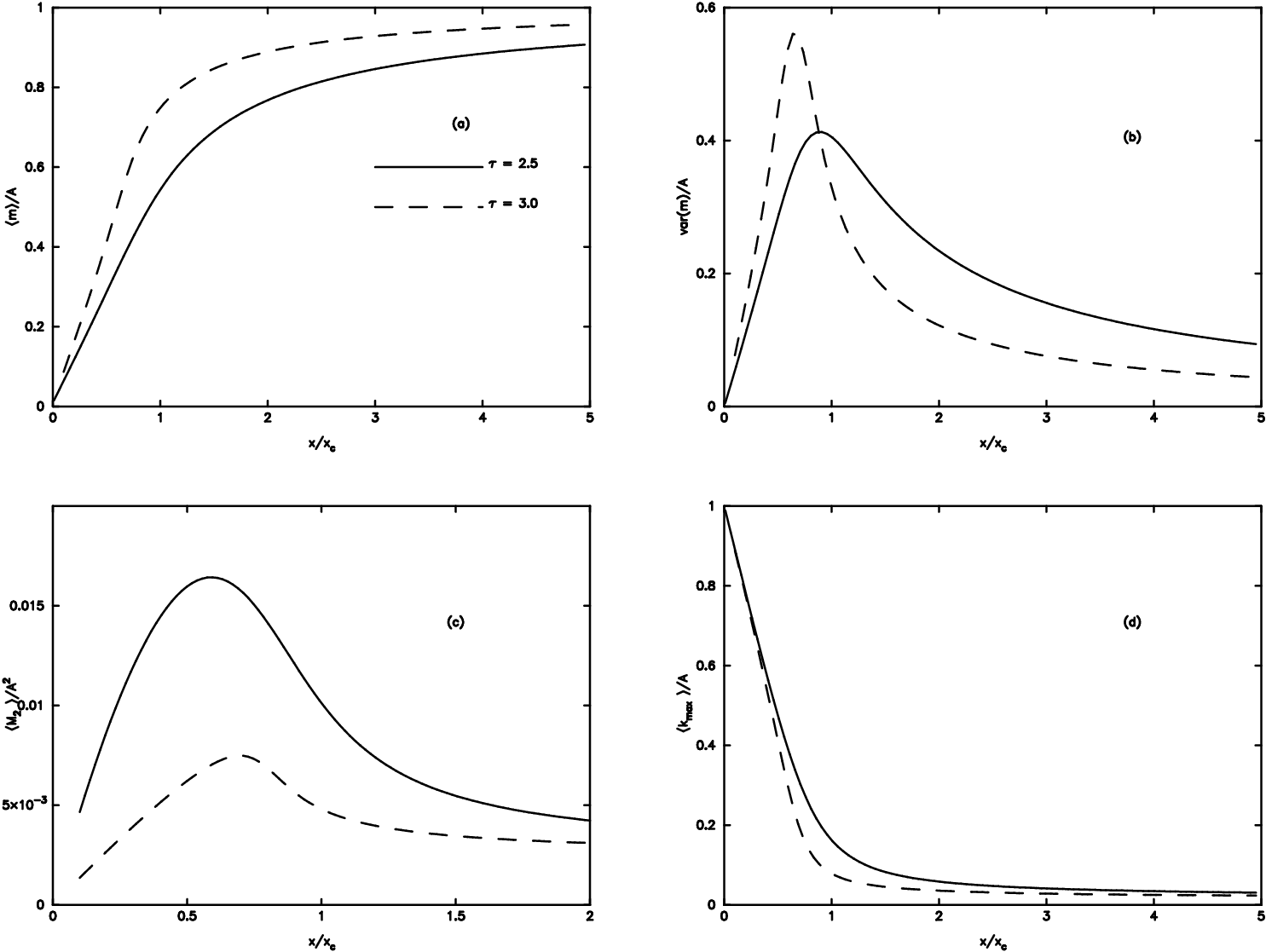}}
\end{figure}

For typical models such as $x_{k} = x/k^{\tau}$ with $\tau > 2$, the
variance of $m$ and the reduced moment $M_{2}$ should peak at the point
$x_{c}/A = 1/\zeta(\tau - 1)$.  The finite size effects can be
considerable, and the $M_{2}$ peak is significantly shifted from its
infinite value in Fig.~\ref{fig:Exp-nuclear}.  At the same point  
$\langle k_{{\rm max}} \rangle$ tends
to zero in the infinite $A$ limit, and tends to be a better indicator
of the transition.  All these facts are consistent with
the interpretation of $x = x_{c}$ as a critical point of a phase
transition characterized by the appearance of an ``infinite'' or
percolating cluster.

\subsection{Disordered Systems}
\label{sec:traditional-disordered}

Disordered systems are traditionally analyzed by considering the
distribution of an observable at a particular point in parameter
space.  The important observables in disordered systems
are $Y=Y_{2}$ and $k_{{\rm max}}$, and their distributions show
significant non-Gaussian behavior, and this feature is in some
sense a characteristic of the disorder.

For the model given by Eq.~(\ref{eq:PM-Pr}), these can be obtained by
a recursive procedure.  Define $x_{k} = (\gamma)_{k-1}/k!$.
Then the probability for a pattern separates into a term like
Eq.~(\ref{eq:Gibbs-Pr}) times a term which depends only on the
multiplicity.  The partition function, $Z_{A} = (x+\gamma)_{A-1}/A!$, 
is then essentially a sum of terms contributed by each multiplicity
class, namely $Z_{A} = \sum_{m} x_{(m-1; 1-\gamma)} Z_{A}^{m}$.
Each of these terms can be computed for any $x_{k}$ from the recursion
$m Z_{A}^{m} = \sum_{k} x_{k} Z_{A-k}^{m-1}$.
Then the probability distribution 
$\Pi(k_{{\rm max}}) = \Delta Z_{A}/Z_{A}$ as in the previous section.

To determine the probability distribution $\Pi(Y)$, the partition
function is broken into 
terms in which $Y$ is fixed, i.e. $Z_{A} = \sum_{Y} Z_{A}^{Y}$, so that
$\Pi(Y) = Z_{A}^{Y}/Z_{A}$.
For computational purposes, it is convenient to continue to break
these partition functions into components where the multiplicity is
fixed, so that  
$Z_{A}^{Y} = \sum_{m} (x)_{(m-1,1-\gamma)} Z_{A}^{m,Y}$.
Using $x_{k}$ as defined above, we can compute $Z_{A}^{m,Y}$ via
the recursion $m Z_{A}^{m,Y} = \sum_{k} x_{k} Z_{A-k}^{m-1,Y-k^{2}}$.

\begin{figure}
\caption{The probability distribution of 
(a) the second moment $\Pi(Y)$ and
(b) the size of the largest cluster $\Pi(k_{\rm max})$
for several disordered systems.}
\label{fig:Pi-disordered}
\centerline{\includegraphics[bb = 126 18 516 744, angle = -90,
            width = 6.0in]{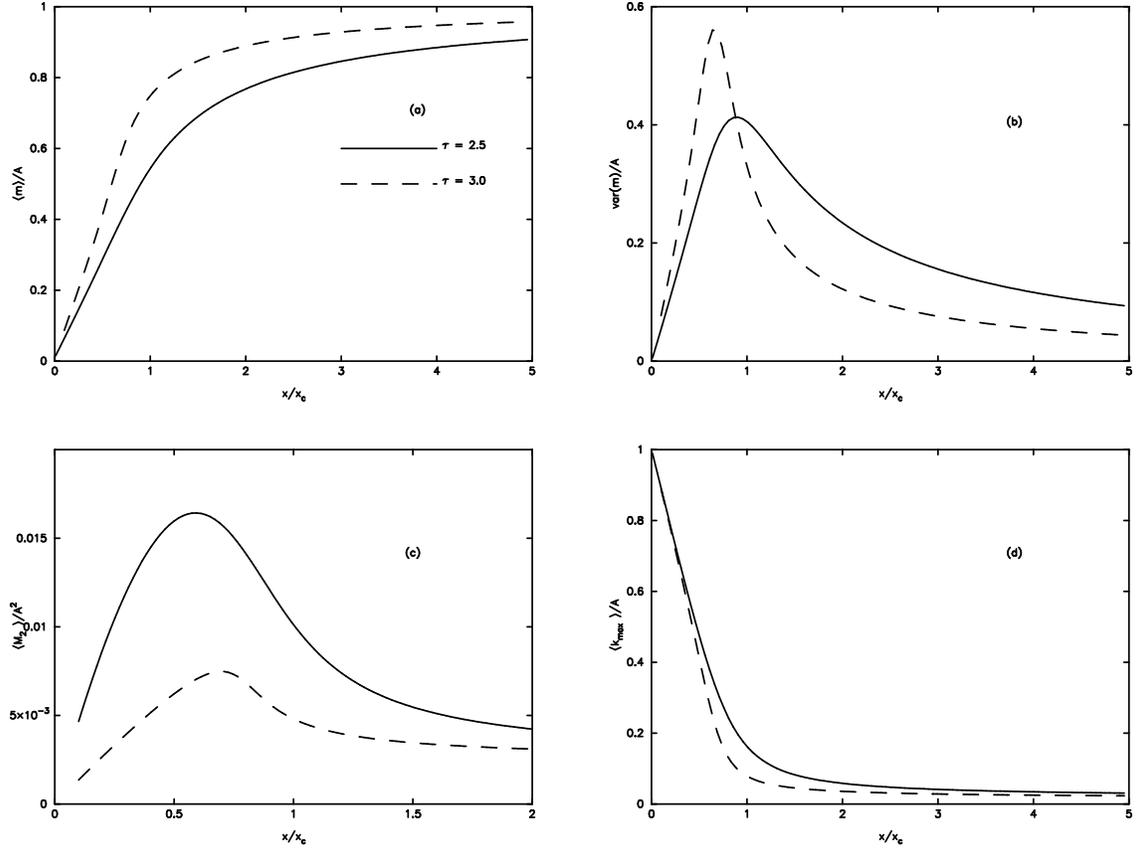}}
\end{figure}

For many values of $x$ and $\gamma$, these distributions are
significantly non-Gaussian, for instance as seen in
Fig.~\ref{fig:Pi-disordered}(a) which shows the distribution
$\Pi({Y})$.
Firstly, one notes the large finite-size fluctuations seen in $\Pi(Y)$
when $Y/A^{2} > 1/2$.  These disappear in the infinite limit.
The presence of cusps in
$\Pi({Y})$ at ${Y/A^{2}} = 1,1/2$ and $1/3$ can be seen in
figure~\ref{fig:Pi-disordered}(a) and persists in the infinite limit.
These cusps are not strictly an indicator of disorder, but the
non-Gaussian nature of the distributions accentuates their presence.

Figure~\ref{fig:Pi-disordered}(b)
which shows $\Pi(k_{\rm max})$ reveals that the distribution of sizes
of the largest cluster varies widely for different models.  
All these cases are likely to have sizable largest clusters, however
there is a nontrivial distribution about $k_{\rm max}/A = 1/2$ for these
models, due in part to the large number of patterns with 
$k_{\rm max} = A/2$.
The cusps at $k_{\rm max}/A =1/2, 1/3$ etc. arise
from singularities associated with breaking the system into fragments
of equal size.  For example, below $k_{\rm max}/A=1/2$, there is no
contribution from binary fragmentation.  The sudden cutoff of this
contribution leads to a cusp.

\section{Corresponding Analysis of Models}
\label{sec:corresponding-analysis}

In this section we reverse our methods of analysis, treating nuclear
models as if they were disordered systems, and vice versa.  Such an
approach is warranted since both descriptions can be described without
reference to the fragmenting mechanisms, and therefore the methods
should be insensitive to the source of the data and sensitive instead
to the underlying property we are trying to isolate in the data
(i.e. criticality and disorder).  Whether the methods are indeed
selective should be revealed by such a (potentially) blind test.

\subsection{Nuclear Systems}
\label{sec:corresponding-nuclear}

The corresponding analysis of fragmenting nuclei considers the
distribution of the second moment $Y$ and the size of the largest
cluster $k_{\rm max}$.
Computing $\Pi(Y)$ and $\Pi(k_{{\rm max}})$ is not complicated for
models with fragmentation pattern given by Eq.~(\ref{eq:Gibbs-Pr}).
We have already discussed how to obtain $\Pi(k_{{\rm max}})$ from the
partition function in section~\ref{sec:traditional-nuclear}.  
The distribution $\Pi(Y)$ is obtained by breaking the contributions to
the partition function $Z_{A}$ into classes with fixed $Y$, 
$Z_{A} = \sum_{Y} Z_{A}^{Y}$ and applying 
$\Pi(Y) = Z_{A}^{Y}/Z_{A}$.
As in section~\ref{sec:traditional-disordered}, these partition
functions can be computed
by recursion, i.e. $A Z_{A}^{Y} = \sum_{k} k x_{k} Z_{A-k}^{Y-k^{2}}$.

\begin{figure}
\caption{The probability distribution of
(a) the second moment $\Pi(Y)$ and
(b) the size of the largest cluster $\Pi(k_{\rm max})$
for the model given by 
Eq.~(\ref{eq:Gibbs-Pr}) with $x_{k} = x/k^{5/2}$ for various $x$.}
\label{fig:Pi-nuclear}
\centerline{\includegraphics[bb = 126 18 516 744, angle = -90,
            width = 6.0in]{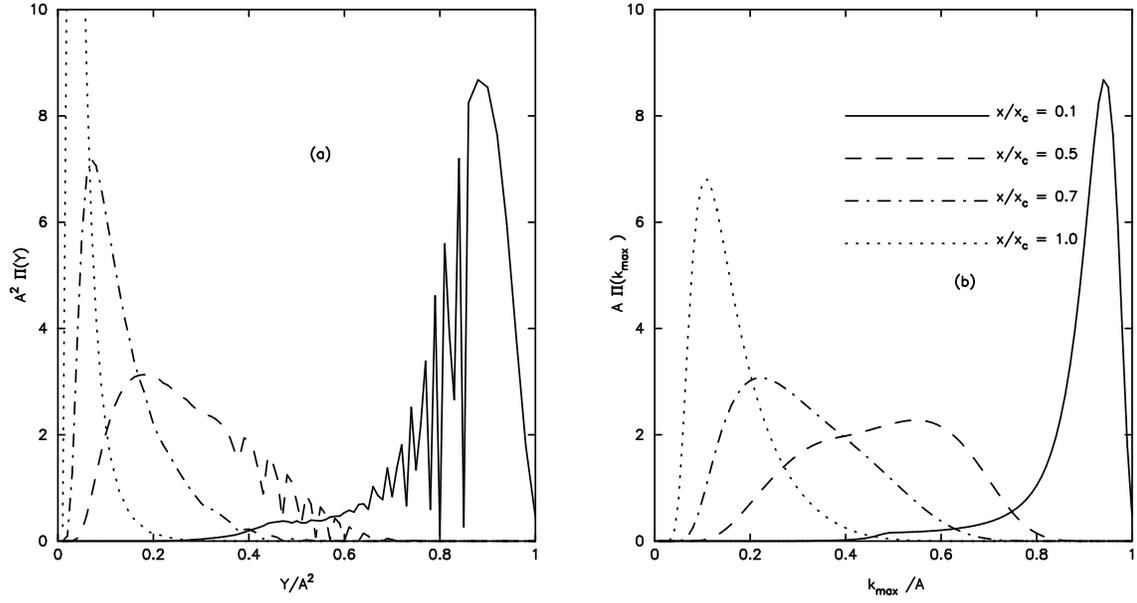}}
\end{figure}

If we analyze the fragmentation distribution given by
Eq.~(\ref{eq:Gibbs-Pr}) with a typical $x_{k} = x/k^{\tau}$ with 
$\tau > 2$ 
by plotting the distribution on the largest cluster size and the
second moment, we find over most of the range that the distributions
are essentially Gaussian centered about their expectation values.
Figure~\ref{fig:Pi-nuclear} shows this.
When $x$ is small and the pattern is usually dominated by a single large
fragment, the atypical events with smaller large fragments 
are distributed in a manner similar to a disordered system with a low
value of $x$, only with a much diminished probability.  This arises
almost entirely from binary fragmentation events.

\subsection{Disordered Systems}
\label{sec:corresponding-disordered}

The corresponding analysis of disordered systems examines the
expectation values of the reduced moments as a function of a changing
parameter.  We consider two cases given by Eq.~(\ref{eq:PM-Pr}): the
randomly broken interval with $\rho(z) = x(1-z)^{x-1}$ with $x>0$
($\gamma = 1$), and the spin-glass model with $0<x<1$ ($x+\gamma=1$).
To facilitate computing the reduced moments for
the models described by Eq.~(\ref{eq:PM-Pr})
we have the following results for ensemble averages of $n_{k}$
\begin{eqnarray}
\langle n_{k} \rangle 
  & = & {A \choose k} 
        {(x)_{A-k} (\gamma)_{k-1} \over (x+\gamma)_{A-1}}
        \label{eq:PM-nk} \\
\langle n_{j}(n_{k} - \delta_{jk}) \rangle
  & = & x {A! \over j! k! (A-j-k)!} (\gamma)_{j-1} (\gamma)_{k-1} 
        \nonumber \\
     && \times   
        {(x-\gamma+1)_{A-j-k} \over (x+\gamma)_{A-1}} \;.
        \label{eq:PM-njnk}
\end{eqnarray}
With these we can compute $\langle Y_{s} \rangle$ and its variance.
Before we do so, it is interesting to note some properties of 
$\langle n_{k} \rangle$ itself.

Since $\sum_{k} k n_{k} = A$, 
$k \langle n_{k} \rangle/A$ is properly a probability distribution.
In this case, the distribution 
has appeared before as a distribution for a replacement urn model
introduced by P\'{o}lya~\cite{Feller-W:1967-intro-prob}.
This correspondence was noted by us before in the special case of
$\gamma = 1$~\cite{Mekjian-AZ:1991pra1}.
A central feature in that case was the appearance of a power law in
the fragment distribution $\langle n_{k} \rangle = 1/k$ for $x=1$.
Such a power law applies in this more 
general case, since in the large $A$ limit, we can approximate
$\langle n_{k} \rangle$ by
\begin{equation}
\langle n_{k} \rangle 
  \approx \frac {\Gamma(x +\gamma)}{\Gamma(x)\Gamma(\gamma)} 
    \frac {1}{k} \left( 1 - \frac {k}{A}\right)^{x-1} 
    \left( \frac {k}{A} \right )^{\gamma-1} \;.
    \label{eq:PM-nk-approx}
\end{equation}
Setting $x = 1$, 
$\langle n_{k} \rangle \approx \gamma A^{1 -
\gamma} k^{\gamma-2}$ which shows a power law falloff with
exponent $\tau = 2 - \gamma$.

If we now allow $A \rightarrow \infty$, $k/A \rightarrow W$, the
discrete model becomes a continuum one.  In this case, 
$f(W) = \lim_{A \rightarrow \infty} A \langle n_{WA} \rangle$
is the continuum distribution of fragment sizes, with 
$\int_{0}^{1} W f(W) dW = 1$ and 
$\lim_{A \rightarrow \infty} \langle Y_{s} \rangle/A^{2} = \int_{0}^{1} W^{s} f(W) dW$.
Using Eq.~(\ref{eq:PM-nk-approx}), we arrive at
\begin{equation}
f(W) 
  = {\Gamma(x+\gamma) \over \Gamma(x) \Gamma(\gamma)}
    (1 - W)^{x-1} W^{\gamma-2} \;.
    \label{eq:beta-distribution}
\end{equation}
i.e., $W f(W)$ is a beta distribution.
Derrida and
Flyvbjerg~\cite{Derrida-B:1986jpa,Derrida-B:1987jpa,Derrida-B:1988jpa},
extending results on spin-glasses due to M\'{e}zard {\em et
al.}~\cite{Mezard-M:1984jdp} and other disordered systems, arrived at
these distributions before.
All the cases they considered are in fact special cases of the above
expression.  Specifically, 
$\gamma = 1, x = 1/2$ reproduces the quenched random map, 
$\gamma = 1, x > 0$ reproduces the randomly broken interval and 
$\gamma = 1 - x$ reproduces the Sherrington-Kirkpatrick spin-glass
model~\cite{Mezard-M:1984jdp}.  We have already seen why the first two
models reduce to this model.  The spin-glass result on the other hand
defies a simple explanation.  Distributions on the largest and
second largest cluster also agree with the known spin-glass results,
and so it appears that this model does indeed capture the clustering
properties of the spin-glass.

Returning now to the question of moments, we see that in this case we
can arrive at closed form solutions.
Ensemble averages for the multiplicity and its variance can be derived
by applying 
Eqs.~(\ref{eq:PM-nk})~and~(\ref{eq:PM-njnk}) and some combinatorial
identities (Eqs.~(\ref{eq:nk-sum-rule})~and~(\ref{eq:njnk-sum-rule}))
to arrive at
\begin{eqnarray}
\langle m \rangle 
   & = & {x+\gamma-1 \over \gamma-1} 
         \left(1 - {(x)_{A} \over (x+\gamma-1)_{A}} \right) \;,
         \\
{\rm Var}(m) 
  & = & {1 \over \gamma -1} 
        {(x)_{A} \over (x+\gamma)_{A-1}}
        \left( 1 - {1 \over \gamma -1} 
        {(x)_{A} \over (x+\gamma)_{A-1}}
        \right.
        \nonumber \\
    && \hspace{0.3in} \left. +{1 \over \gamma-1} 
        {(x-\gamma+1)_{A} \over (x+1)_{A-1}} \right) \;,
\end{eqnarray}
Similarly, the mean and variance of $Y$ can be determined
\begin{eqnarray}
\langle {Y} \rangle 
  & = & \frac {\gamma}{x + \gamma} \left ( 1+ \frac{1}{A}
        \frac {x}{\gamma} \right)
        \label{eq:PM-expY}
        \\
{\rm Var}(Y)
  & = & {2 x \gamma (A-1) (x+\gamma+A)(x+\gamma+A-1) \over 
        A^{3} (x+\gamma)^2 (x+\gamma+1)(x+\gamma+2)} \;.
       \label{eq:PM-varY}
\end{eqnarray}
From this the relative fluctuations in $Y$, $f(Y) = \langle Y^{2}
\rangle/\langle Y \rangle^{2} - 1$ can be computed.  They are
important when determining whether a system is
self-averaging~\cite{Mezard-M:1987-SGTaB}. Consider the case
$\gamma=1$,
where $f(Y) \approx 2x/((x+2)(x+3))$ in the large $A$ limit. For $x
\approx 1$, the fluctuations are large and the system is said to lack
self-averaging properties. For $x \propto A$ the fluctuations are small
and $f(Y) \sim 1/A$ as expected. This is true in general in this
model, and for the spin glass case $x + \gamma = 1$ leads to the known
result $\langle Y^{2} \rangle = 1/3 \langle Y \rangle (1 + 2 \langle Y
\rangle)$~\cite{Mezard-M:1984jdp} in the infinite $A$ limit.


The self-averaging property is not difficult to understand.
At one extreme, the nucleus can be broken into individual nucleons
so that ${Y/A^2} = 1/A$; thus, $Y$ varies inversely as the size of
the system.  At the other extreme, when the $A$ nucleons are in
one cluster ${Y/A^2} = 1$.  This
behavior is analogous to the two extreme limits for the participation
ratio in Anderson localization. For a localized state ${Y/A^2} \sim 1/A$
and for an extended state ${Y/A^2} \sim 1$. A localized state corresponds
to the localization of the electron wave functions around many
individual sites, while an extended state is the opposite case. A
percolation description of localized versus extended states can be
found in Ziman~\cite{Ziman-JM:1979-MoD}. Specifically, for site
percolation an extended state includes an incipient infinite cluster.

\begin{figure}
\caption{The expected value (a) and variance (b) of the multiplicity
and the expected values of the reduced second moment (c) and
the largest cluster (d) for the 
random broken interval (solid line) and
spin-glass (dashed line) models.
Note that the spin glass is only defined when $0<x<1$.}
\label{fig:Exp-disordered}
\centerline{\includegraphics[bb = 43 20 592 746, angle = -90,
            width = 6.0in]{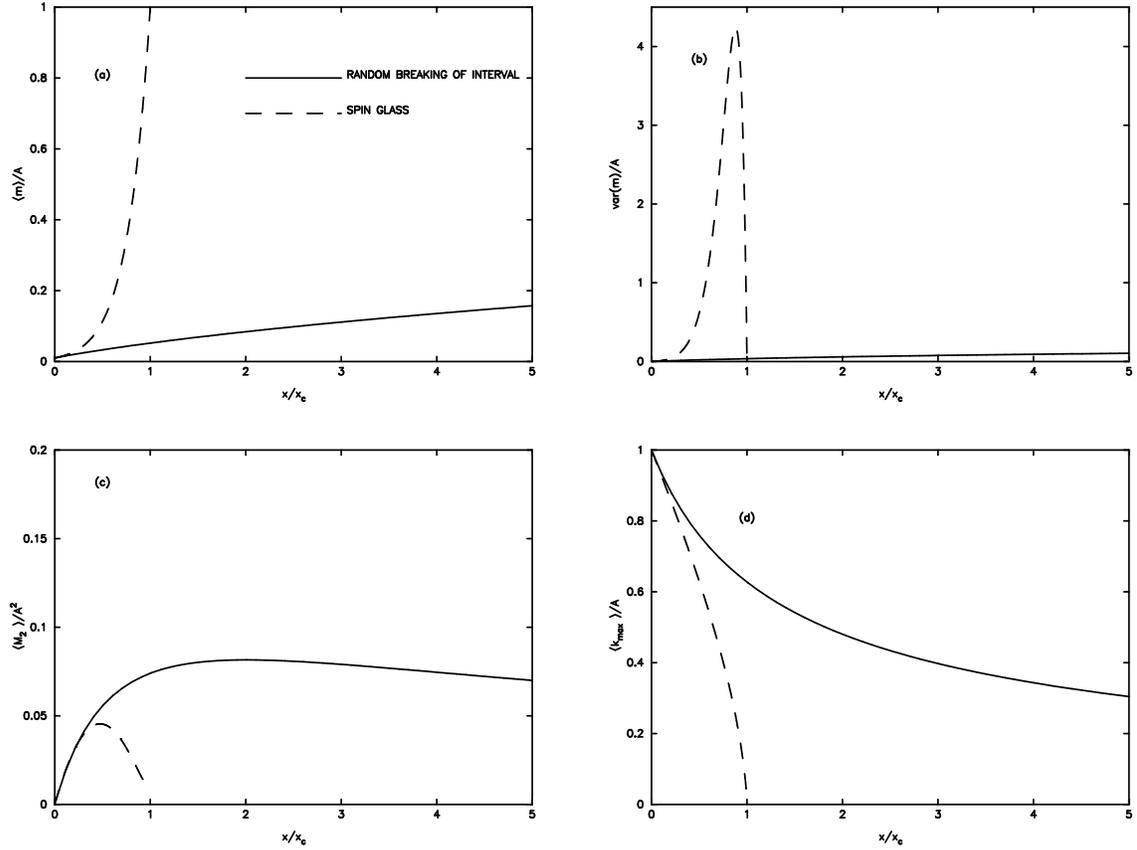}}
\end{figure}

Applying these results, we plot the various expectation values in 
Fig.~\ref{fig:Exp-disordered}.  From these graphs one can comment
on the possibility of criticality in this model.  It is well known
that the $x>0$, $\gamma=1$ case is not critical in the traditional
sense.  The zeros of the partition function occur on the negative real
$x$ axis, well isolated from the relevant parameter domain.
For the $x+\gamma = 1$, $0 < x < 1$ model,
the reduced moments appear to peak near $x = 1/2$,
which suggests that $x=1/2$ may be a critical point for the model.
However, ${\rm Var}(m)$ peaks closer to $x = 0.9$.  At a true critical
point one would expect this fluctuation to be maximal at the critical
point.  Since for any model the various expectation values must peak
somewhere, the existence of a maximum is insufficient evidence for
criticality.  
Further evidence is inferred from the fact that the partition function
$(x+\gamma)_{A-1}/A!$ has zeroes on the negative real $x+\gamma$ axis in
a complexified parameter space.  These zeroes are isolated from
$x+\gamma = 1$ real space, and suggests that a traditional critical point
in which the zeros of the partition function encroach the
thermodynamically relevant parameter domain is completely absent in
this model.

\section{Conclusions and Outlook}
\label{sec:conclusion}


In this paper we have shown that the random breaking of 
an interval process when suitably generalized
encompasses a wide range of physical
models both in nuclear fragmentation and in disordered systems. 
The fact that all these models are implicit to a particular process
strongly recommends that the analysis of their properties should be
the same.  Traditionally however this has not been the case. The models
have arisen separately, motivated by particular features which
theorists wished to capture. In nuclear fragmentation, the feature to
reproduce is a specific criticality marked by the appearance of an
``infinite'' cluster.  Interest has therefore focussed on reduced
moments whose behavior changes with the appearance of
such a cluster. 
In disordered systems the interest instead has been on
whether such systems lack self-averaging properties,
i.e. that have significant non-Gaussian fluctuations.  
In this case, the focus is on
probability distributions of the measures which might express such
non-Gaussian fluctuations, e.g. the largest cluster and the second
moment.  

Given then this separate emphasis, it
is understood why the same variables in the models are studied 
in different ways.  The corresponding analysis taken here
reinforces the distinct approaches, as it shows that the 
general behavior of the two classes of models differ 
considerably.  Nevertheless, the convergence of their description
within the random breaking of an interval mechanism makes such a
distinction somewhat puzzling.  This mechanism is unlikely to
be physical in the nuclear case, but this is not an explanation for
the different behavior in the two cases.  Indeed, the explanation may
have a great deal to do with criticality or its absence.  Models with
exponent $\tau<2$ tend to look disordered, $\tau>2$ tend to have distinct
critical points and Gaussian distributions.  Since $\tau = 2 +
1/\delta$, the two regions are characterized by positive and negative
$\delta$.  It would be interesting to confirm this observation as a
general feature.

\acknowledgements

This work was sponsored in part by DOE grant \#DE-F602-96ER40987.
We'd like to thank H. Neuberger for some very useful discussions.

\appendix

\section{A combinatorial identity}
\label{sec:perm-identity}

Recall that a permutation on $n$ elements can be described by the
cycles containing the elements which permute among themselves under
the iterated action of the permutation.  For example, 
the permutation $p = \left({1 2 3 4 5 6 7 8 \atop 3 1 2 5 4 6 8 7} \right)$
is usually written in cycle notation as $(132)(45)(6)(78)$.
If each cycle of length $k$ is associated with a cluster of the size
$k$, and these clusters are ordered by the
largest element in each cluster, then each permutation can be
mapped to a sequential partitioning.  The above example
would correspond to the sequential partition $(2,1,2,3)$.

Suppose one chooses a permutation $p \in S_{n}$ uniformly at random,
i.e. with probability $1/n!$.  What is the probability that the
first cluster in the sequence has size $\lambda_{1}=k$?   There are 
${n-1 \choose k-1}$ ways of choosing the $k-1$ elements which are not
element $n$ but are in $n$'s cycle map.  There are $(k-1)!$ different
cycles which permute the 
$k$ objects and $(n-k)!$ ways of permuting the remaining elements
which are not in this cycle.  So the probability is 
$(k-1)! (n-k)! {n-1 \choose k-1}/n! = 1/n$.
What about the second cluster?  The $(n-k)!$ ways of permuting the
elements not in the first cluster is a 
permutation $p' \in S_{n-\lambda_{1}}$ chosen uniformly at random.  In
this permutation, 
$\lambda_{2}$ is the first element, and by the above argument has size
$\lambda_{2} = k$ with probability $1/(n-\lambda_{1})$.  Iterating yields
the probability of the whole sequence as being
$\Pr(\lambda_{1}, \ldots, \lambda_{m}) = 1/
(n(n-\lambda_{1})(n-\lambda_{1}-\lambda_{2}) \cdots)$.
Summing this over all possible orders one obtains
$1/n!$ times the number of permutations with this particular cycle
class structure.  It is well known that this is given by Cauchy's
formula~\cite{Riordan-J:1958-intro-combin}
$n!/\prod_{k} n_{k}! k^{n_{k}}$, so that
\begin{eqnarray}
\sum_{{\lambda} \mapsto {\mathbf{n}}} 
        {1 \over n(n - \lambda_{1})(n-\lambda_{1}-\lambda_{2}) \cdots}
  & = & \prod_{k} {1 \over n_{k}!} 
        \left ({1 \over k} \right )^{n_{k}}
        \label{eq:perm-identity}
\end{eqnarray}

\section{Sum rules}
\label{sec:sum-rules}

For systems satisfying
Eqs.~(\ref{eq:PM-nk})~and~(\ref{eq:PM-njnk}) one has the
following sum rules
\begin{eqnarray}
S_{p} 
  & = & \sum_{k=p}^{A} {[k]_{p} \over [A]_{p}}
        \langle n_{k} \rangle 
    =   {(\gamma)_{p-1} \over (x+\gamma)_{p-1}} \;,
        \label{eq:nk-sum-rule}
        \\
S_{pq}
  & = & \sum_{jk} {[j]_{p} [k]_{q} \over [A]_{p+q}} 
        \langle n_{j} (n_{k} - \delta_{jk}) \rangle
    =   x {(\gamma)_{p-1} (\gamma)_{q-1} \over  (x+\gamma)_{p+q-1}} \;,
        \label{eq:njnk-sum-rule}
\end{eqnarray}
where $p,q>0$, $[k]_{0} = 1$, $[k]_{p} = k(k-1) \ldots (k-p+1)$.
Notice that the right hand sides are independent of $A$.
These rules are helpful in determining expectation values of $m, Y$
etc. and can be derived using a common combinatorial identity known as
Norl{\"u}nd's formula~\cite{Tomescu-I:1975-intro-combin}.


\begin{thebibliography}{10}

\bibitem{Kreutz-P:1993npa}
P. Kreutz {\it et~al.}, Nucl. Phys. A {\bf 556},  672  (1993).

\bibitem{Hauger-JA:1996prl}
J.~A. Hauger {\it et~al.}, Phys. Rev. Lett. {\bf 77},  235  (1996).

\bibitem{Pochodzalla-J:1995prl}
J. Pochodzalla {\it et~al.}, Phys. Rev. Lett. {\bf 75},  1040  (1995), {ALADIN}
  Collaboration.

\bibitem{Wang-S:1995prl}
S. Wang {\it et~al.}, Phys. Rev. Lett. {\bf 74},  2646  (1995), {EOS}
  Collaboration.

\bibitem{Gilkes-ML:1994prl}
M.~L. Gilkes {\it et~al.}, Phys. Rev. Lett. {\bf 72},  1590  (1994), {EOS}
  Collaboration.

\bibitem{Gross-DHE:1990rpp}
D.~H.~E. Gross, Rep. Prog. Phys. {\bf 53},  605  (1990).

\bibitem{Bondorf-JP:1995prep}
J.~P. Bondorf {\it et~al.}, Phys. Rep. {\bf 257},  133  (1995).

\bibitem{Pan-JC:1995prc}
J. Pan and S. Das~Gupta, Phys. Rev. C {\bf 51},  1384  (1995).

\bibitem{Elliott-JB:1994prc}
J.~B. Elliott {\it et~al.}, Phys. Rev. C {\bf 49},  3185  (1994), {EOS}
  Collaboration.

\bibitem{Campi-X:1986jpa}
X. Campi, J. Phys. A {\bf 19},  L917  (1986).

\bibitem{Derrida-B:1986jpa}
B. Derrida and H. Flyvbjerg, J. Phys. A {\bf 19},  L1003  (1987).

\bibitem{Ziman-JM:1979-MoD}
J.~M. Ziman, {\em Models of Disorder} (Cambridge Univ. Press, Cambridge, 1979).

\bibitem{Mezard-M:1987-SGTaB}
M. M\'{e}zard, G. Parisi, and M.~A. Virasoro, {\em Spin Glass Theory and
  Beyond}, Vol.~9 of {\em Lecture Notes in Physics} (World Scientific,
  Singapore, 1987).

\bibitem{Chowdhury-D:1986-SGFS}
D. Chowdhury, {\em Spin Glasses and other Frustrated Systems} (Princeton Univ.
  Press, Princeton, NJ, 1986).

\bibitem{Higgs-PG:1995pre}
P.~G. Higgs, Phys. Rev. E {\bf 51},  95  (1995).

\bibitem{Bauer-W:1985plb}
W. Bauer, D. Dean, U. Mosel, and U. Post, Phys. Lett. B {\bf 150},  53  (1985).

\bibitem{Mekjian-AZ:1997pla}
A.~Z. Mekjian and K.~C. Chase, Phys. Lett. A {\bf 229},  340  (1997).

\bibitem{Anderson-PW:1958prev}
P.~W. Anderson, Phys. Rev. {\bf 109},  1492  (1958).

\bibitem{Sobotka-LG:1985prc}
L. Sobotka and L. Moretto, Phys. Rev. C {\bf 31},  668  (1985).

\bibitem{Chase-KC:1995prl}
K.~C. Chase and A.~Z. Mekjian, Phys. Rev. Lett. {\bf 75},  4732  (1995).

\bibitem{Chase-KC:1994prc1}
K.~C. Chase and A.~Z. Mekjian, Phys. Rev. C {\bf 49},  2164  (1994).

\bibitem{Lee-SJ:1992prc2}
S.~J. Lee and A.~Z. Mekjian, Phys. Rev. C {\bf 45},  1284  (1992).

\bibitem{Fisher-ME:1967phy}
M.~E. Fisher, Physics (N.Y.) {\bf 3},  255  (1967).

\bibitem{Gross-DHE:1992adp1}
D.~H.~E. Gross, A. Ecker, and A.~R. DeAngelis, Ann. Phys. (Leipzig) {\bf 1},
  340  (1992).

\bibitem{Gross-DHE:1992npa}
D.~H.~E. Gross, A. Ecker, and A.~R. DeAngelis, Nucl. Phys. A {\bf 545},  187c
  (1992).

\bibitem{Jaqaman-HR:1992npa}
H.~R. Jaqaman, A.~R. DeAngelis, A. Ecker, and D.~H.~E. Gross, Nucl. Phys. A
  {\bf 541},  492  (1992).

\bibitem{Mekjian-AZ:1990prl}
A.~Z. Mekjian, Phys. Rev. Lett. {\bf 64},  2125  (1990).

\bibitem{Derrida-B:1987jdp}
B. Derrida and H. Flyvbjerg, J. de Physique {\bf 48},  971  (1987).

\bibitem{Derrida-B:1987jpa}
B. Derrida and H. Flyvbjerg, J. Phys. A {\bf 20},  5273  (1987).

\bibitem{Frontera-C:1995pre}
C. Frontera, J. Goicoechea, I. R\`{a}olfs, and E. Vives, Phys. Rev. E {\bf 52},
   5671  (1995).

\bibitem{Nijenhuis-A:1978-combin-algorithms}
A. Nijenhuis and H.~S. Wilf, {\em Combinatorial Algorithms}, {\em Computer
  Science and Applied Mathematics}, 2nd ed. (Academic Press, Orlando, FL,
  1978).

\bibitem{Donnelly-PJ:1991aap}
P.~J. Donnelly, W.~J. Ewens, and S. Padmadisastra, Adv. Appl. Prob. {\bf 23},
  437  (1991).

\bibitem{Derrida-B:1988jpa}
B. Derrida and D. Bessis, J. Phys. A {\bf 21},  L509  (1988).

\bibitem{Pitman-J:1995ptrf}
J. Pitman, Prob. Theory. Relat. Fields {\bf 102},  145  (1995).

\bibitem{Sherrington-D:1975prl}
D. Sherrington and S. Kirklpatrick, Phys. Rev. Lett. {\bf 35},  1792  (1975).

\bibitem{Derrida-B:1980prl}
B. Derrida, Phys. Rev. Lett. {\bf 45},  79  (1980).

\bibitem{Campi-X:1992npa}
X. Campi and H. Krivine, Nucl. Phys. A {\bf 545},  161c  (1992).

\bibitem{Chase-KC:1995prc}
K.~C. Chase and A.~Z. Mekjian, Phys. Rev. C {\bf 52},  R2339  (1995).

\bibitem{Feller-W:1967-intro-prob}
W. Feller, {\em An Introduction to Probability Theory}, 3rd ed. (John Wiley \&
  Sons, New York, 1967).

\bibitem{Mekjian-AZ:1991pra1}
A.~Z. Mekjian and S.~J. Lee, Phys. Rev. A {\bf 44},  6294  (1991).

\bibitem{Mezard-M:1984jdp}
M. M\'{e}zard {\it et~al.}, J. de Physique {\bf 45},  843  (1984).

\bibitem{Riordan-J:1958-intro-combin}
J. Riordan, {\em An Introduction to Combinatorial Analysis} (John Wiley \&
  Sons, New York, 1958).

\bibitem{Tomescu-I:1975-intro-combin}
I. Tomescu and S. Rudeanu, {\em Introduction to Combinatorics} (Collet's,
  London, 1975).

\end{thebibliography}

\hsize\textwidth\columnwidth\hsize\csname @twocolumnfalse\endcsname
\widetext

\end{document}